# Automated Construction of Artificial Lattice Structures with Designer Electronic States

Ganesh Narasimha[1†], Mykola Telychko[1†], Wooin Yang[1], Arthur P. Baddorf[1], P. Ganesh[1], An-Ping Li[1]*, Rama Vasudevan[1]*

[1]*Center for Nanophase Materials Sciences (CNMS), Oak Ridge National Laboratory (ORNL), Oak Ridge, Tennessee, USA – 37831*

* apli@ornl.gov, vasudevanrk@ornl.gov

**Abstract**

Manipulating matter with a scanning tunneling microscope (STM) enables creation of atomically defined artificial structures that host designer quantum states. However, the time-consuming nature of the manipulation process, coupled with the sensitivity of the STM tip, constrains the exploration of diverse configurations and limits the size of designed features. In this study, we present a reinforcement learning (RL)-based framework for creating artificial structures by spatially manipulating carbon monoxide (CO) molecules on a copper substrate using the STM tip. The automated workflow combines molecule detection and manipulation, employing deep learning-based object detection to locate CO molecules and linear assignment algorithms to allocate these molecules to designated target sites. We initially perform molecule maneuvering based on randomized parameter sampling for sample bias, tunneling current setpoint and manipulation speed. This dataset is then structured into an action trajectory used to train an RL agent. The model is subsequently deployed on the STM for real-time fine-tuning of manipulation parameters during structure construction. Our approach incorporates path planning protocols coupled with active drift compensation to enable atomically precise fabrication of structures with significantly reduced human input while realizing larger-scale artificial lattices with desired electronic properties. Using our approach, we demonstrate the automated construction of an extended artificial graphene lattice and confirm the existence of characteristic Dirac point in its electronic structure. Further challenges to RL-based structural assembly scalability are discussed.

## Introduction

The pursuit of matter manipulation at the fundamental level is driven by the quest to design materials with tailored properties. Since its first demonstration[1], the scanning tunneling microscope (STM) has been a reliable tool to create atomic scale structures that host designer states. For example, arranging atoms or molecules in specific geometries exhibits 1D and 2D electronic bands, different from the pristine substrate.[2, 3] This procedure is also used to modulate electronic properties of 2D materials by controlled manipulation of dopants and defects.[4-7] In addition to physical manipulation, the STM tip can be used to induce chemical reactions create precise heterostructures[8, 9] and emulate molecular structures[10]. Elaborate manipulation procedures have demonstrated miniature functional systems such as a p-n junction,[11] logic gates,[9, 12] a Boltzmann machine,[13] and memory devices.[14]

The application of atomic-scale manipulation with STM relies on its ability to observe and influence structures with atomic precision. Precise manipulation using the STM requires optimizing the conditions related to tip-molecule interactions. Parameters such as molecule position, tip-sample bias, tunneling current setpoint and the speed of the manipulation affects the success of manipulation.[15, 16] A wrong choice of the parameters could either result in an unsuccessful event, or worse, result in a changes of tip apex – rendering the tip ineffective. In such a scenario, the operator undertakes a time-consuming treatment procedure for tip restoration before reinitiating the manipulation process. Further, to host designer states, creation of multi-atom/molecular structures requires stable tip conditions over long periods of the manipulation process. This procedure, which invariably requires close approach conditions, induces perturbations to the tip. The sensitive nature of the tip-adatom interactions, combined with the laborious process of manipulation, limits the exploration of diverse artificial structures.

In recent years, machine learning has gained traction within the scanning probe community to design workflows that improve experimental efficiency.[17-19] Among these methods, Deep Reinforcement Learning (RL) is particularly suited to navigate dynamic environments.[20] RL leverages deep neural networks to enable agents to learn optimal decision-making through iterative interactions with the environment. In the STM experimentation, automation methods were previously implemented by combining elements of image processing and path planning methods with instrumentation controls.[21, 22] Most prominently, I-Ju et.al demonstrated RL for horizontal manipulation of Ag atoms for the creation of artificial structures.[23] In addition to this, other groups have reported RL methods for molecular manipulation, tip-induced surface reactions, and molecular assembly.[24-27] However, fully automated fabrication of artificial lattices using STM has not been demonstrated to date.

In this work, we demonstrate the utilization of RL for semi-automated construction of an artificial lattice by manipulating carbon-monoxide (CO) molecules on Cu (111) surface. The Cu (111) exhibits an electronic surface state which is both approximately two-dimensional and free electron-like. Adsorbed CO ties up these surface state electrons locally, i.e. at the adsorption site and nearest neighbors.[28] Local charge transfer between CO and Cu(111) perturbs the otherwise nearly free-electron like metallic surface states and different assemblies of CO leads to a variety of standing wave patterns due to interference of these oscillations. Selective placement of CO on Cu(111) can therefore create a corresponding geometric pattern of free-electron states, even enabling emergence of artificial lattices with tailored electronic structure. This merit of CO-decorated Cu(111) has been explored to realize exotic configurations such as artificial electronic lattices, quasicrystals and fractal structures.[3, 29, 30] This combination of the CO molecules assembled on the Cu surface provides an ideal testbed to test the efficacy of designing artificial structures that exhibit unique electronic signatures. This work demonstrates the utility of automated manipulation procedure to construct an artificial lattice structure that hosts designer electronic states.

Within the automated workflow, the RL method is used to learn the optimal parameters that results in a higher probability of successful manipulation. The RL method uses an agent (model) that interacts with the environment (STM and CO decorated Cu(111) surface) - observes the current state of the system, decides on the action (manipulation) and receives a reward. This process continues until termination – indicating end of task. The sequence of state, action, reward and termination is then used to generate a policy – a strategy to choose the best action in a given state. In contrast to prior works[23-26, 31], where the agent interacts dynamically with the environment and optimizes the decision based on trial and error processes, we initially employed an offline training approach.[32] Here, the manipulations are carried out across the parameter space, and the data is formulated into a sequence of state, action, reward, and termination, which is then utilized to train the RL agent. This method allows us to be conservative while working with smaller datasets and allows us to implement procedures such as hyperparameter optimization, data augmentation, and narrowed parameter search based on expert knowledge. Once trained, the model is deployed on the STM and fine-tuned through experimental operations during the creation of the artificial structures.

## Results

Our experiments utilize an automated workflow that integrates the manipulation routines, image scanning, data collection, and analysis. These were achieved using python codes that communicate with the STM using the *Nanonis-TCP* port.[33] **Figure 1** illustrates the schematic of the experimental workflow – here, we initially perform manipulation experiments that are then used to train the RL agent, followed by deployment on the STM for fine-tuning of manipulation parameters. The workflow begins with STM imaging to capture the initial state of the system as represented by **Figure 1a,** which shows the CO molecules dispersed on the Cu (111) surface. These images are used to resolve the position of the CO molecules using a convolutional neural network (CNN) based You-Only-Look-Once (YOLO) object detection method.[34] **Figure 1b** shows the model detections of CO molecules represented by the bounding boxes. Prior to deploying the YOLO-model into the framework, we first trained on numerous annotated images to detect the CO molecules.

Next, we randomly fix a target within the scan area. A molecule is designated to the target position (as shown in **Figure 1c**) using a linear assignment algorithm based on distance minimization. This method is scalable to multimolecular configurations to retain one-to-one molecule-to-target assignments.

**Figure 1d** demonstrates the manipulation procedure using the STM tip. The manipulation iteration involves positioning the tip at the CO molecule location, lowering the tip height to be in close proximity to the molecule, and then moving to the target position. Here, the manipulation parameters – sample bias ($V_s$), setpoint current ($I_t$), and manipulation speed ($S$) influence the success of manipulation and are designated as the action space of the RL model. The action parameters are determined by the RL policy. To collect the training set, we initially perform manipulation procedures by action parameters based on a random sampling within a given range of $V_S$: [10, 100] mV; $I_t$: [10, 90] nA, and $S$: [0.5, 5] nm/s. Given fewer "successful" events in this larger parameter range, we augment the dataset with further experimental datasets acquired with parameter sampling based on expert knowledge of a lower $V_S$ range: [10, 30] mV and a higher $I_t$ range: [70, 90] nA. During the construction of a structure, the random policy is replaced by the RL agent policy that predicts the manipulation parameters.

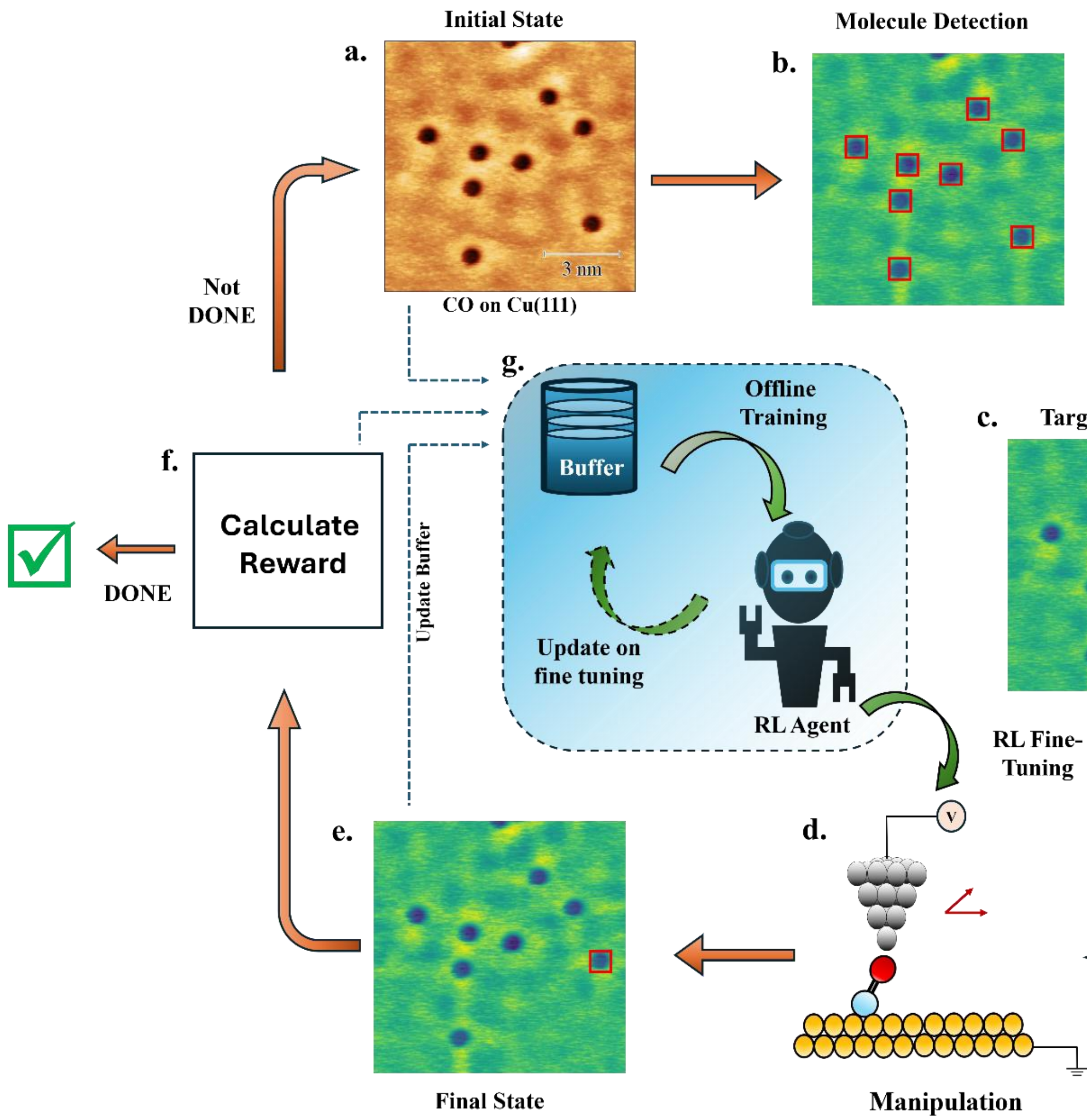

**Figure 1**: Schematic of the autonomous workflow for manipulating CO molecules on a Cu (111) surface. **(a)** The STM scanned image of the CO molecules on the Cu (111) surface. This is used to determine the initial state of the system. **(b)** YOLO object detection is used to detect the molecules and their position. **(c)** A target is fixed, and a molecule is designated to the target based on linear assignment. **(d)** The next step involves the automated manipulation sequence. The manipulation parameters are determined based on a sampling policy. **(e)** The final state is determined after the manipulation procedure. **(f)** The reward is calculated at the end of the manipulation procedure. The procedure is complete if the manipulation event is "DONE". If not, the previous state is iteratively considered as the next initial state. **(g)** The RL procedures

initially involve offline training using the buffer of randomly sampled action parameters and the target. Once trained, the RL agent is deployed on the microscope for fine-tuning of manipulation parameters.

**Figure 1e** shows the final state of the system configuration after the manipulation process. Depending on the molecule distance to the target position, a reward is calculated to evaluate the success of the manipulation (**Figure 1f**). The success of the manipulation is determined by the reward function, which depends on the initial position and current position of the molecule with respect to the target.

$$Reward = \begin{cases} 10 * \left(1 - \frac{|X_{current} - X_{target}|}{d_0}\right), & not\ DONE\ if\ |X_{current} - X_{target}| \geq \varepsilon \\ 100, & \boldsymbol{DONE}\ if\ |X_{current} - X_{target}| < \varepsilon \end{cases}$$

Where $X_{current}$ and $X_{target}$, and $d_0$ are the current molecule position, target position, and the initial manipulation distance, respectively. The above equation indicates a linearly increasing reward as the molecule moves towards the target. Once the molecule reaches the target, we assign an asymmetrically high reward = 100. The consideration of a successful manipulation depends on a tolerance parameter given by $\varepsilon$. A manipulation event is considered "DONE" or successful when the final position of the CO molecule is within a distance of $\varepsilon$ from the target. The transitions associated with manipulation are formulated into an action trajectory that are segmented into episodes. Each episode lasts for 10 iterations or until the manipulation is successful, whichever is earlier.

The manipulation procedures are formulated into an MDP (Markov Decision Process) dataset that are enlisted as state, action, reward, and termination for every iteration. We use this buffer for offline training of the RL model as shown in **Figure 1g**. We use the *d3rlpy* package as the code base for creating and training the RL agent model.[35] We used a Deep Deterministic Policy Gradient (DDPG) algorithm of the RL model to train the model based on the buffer data. DDPG is a lightweight and sample-efficient algorithm that is effective in continuous action spaces, though its convergence can be sensitive to hyperparameter settings.[36, 37] Given that each of the manipulation events accompanied by scanning procedure is a time-consuming process, we are limited by the small dataset for efficient training of the model. Typical RL settings employ $10^4$ - $10^6$ samples to achieve stable policy convergence. In contrast, our experiments utilized 622 episodes, corresponding to ~5000 manipulation events. The sample size is comparable to previous report on RL based STM-manipulation.[23] Operating in the low-sample regime required tuning the hyperparameters accordingly. The offline training methodology enabled systematic hyperparameter optimization by permitting repeated evaluation and refinement without additional experimental data. It also facilitated controlled data augmentation, wherein we shuffled the episodes and introduced Gaussian noise of varying magnitudes to the state variables thereby improving model generalization. Moreover, it is to be noted that workflow supports online fine-tuning for real-time weight updates from real-time experimentation.

The RL model consists of two sets of neural networks – one termed the "actor" and the other termed the "critic". The actor network determines the policy for optimal mapping of the state to the action while the critic network evaluates the policy of the actor network. During model training, the loss of both the critic and the actor were found to reduce globally (as shown in **Figure S1**).We found stable training at lower learning rate (1e-6) extended over longer epochs. However, we see that after the initial five steps (each step indicates 2000 model updates), the critic loss increases, indicating a training imbalance caused by erroneous value function estimation. This suggests that the policy is being optimized based on inaccurate value estimates, potentially leading to overfitting on the training data and poor generalization to new samples. Therefore, we chose the model trained at step five to be deployed on the microscope for online fine-tuning of manipulation parameters.

As pointed out above, the process of manipulation involves approaching the STM tip to the molecule, followed by movement to the target position. A schematic of this is illustrated in **Figure 2a**. During the course of the manipulation events, we encountered situations wherein the molecules tend to become stuck at certain locations - either at Cu lattice sites or they tend to agglomerate with other CO molecules. To address these issues, we introduced path-planning algorithms to facilitate the movement of these molecules with a suitable algorithm to detect situations when the molecules could become immobile. The first modification is the introduction of an offset to the manipulation with respect to the target position, which is shown in the schematic **Figure 2b**. This is particularly suited to moving small distances (< 1 nm) and to disentangle joint CO molecules.

The second strategy involves optimizing the manipulation path along the high-symmetry crystallographic axes of Cu(111) surface. A prerequisite for this procedure is the aligned orientation of the underlying Cu lattice with respect to the structure that we intend to construct. We initially determined the Cu-lattice orientation by arranging the CO molecules in a known configuration, as shown in **Figure 2c** (further details in **Figure S6**). Since the CO molecules tend to occupy "on-top" site positions over Cu(111) lattice, we designed an algorithm to set the manipulation angle so as to align the path along the Cu-axis, as illustrated in **Figure 2d** (Further details in **Figure S10** of Methods section). This methodology combining path planning with the RL agent is in contrast with previous works, which did not use the low energy pathways afforded by the crystal orientations.

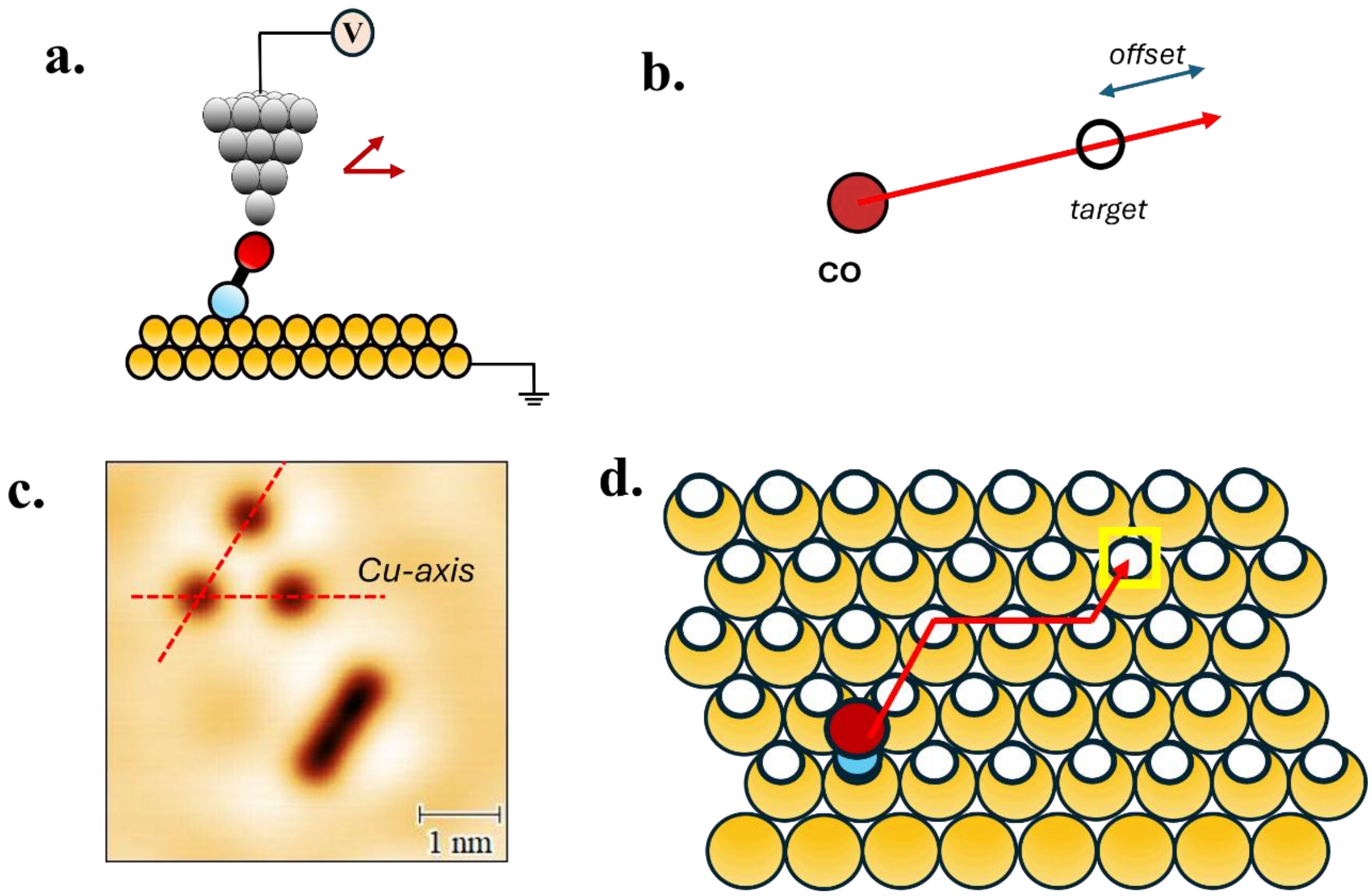

**Figure 2**: Path planning methods used in the STM manipulation routine. **(a)** Schematic of the STM manipulation. **(b)** Illustration of the target offset method for molecule positioning. **(c)** Determination of the Cu-lattice orientation using the CO molecules. The distance between the disjoint molecules is *4a* (*a* being the lattice constant of Cu (111) surface), while that of the joint molecules is *2a*. **(d)** Depiction of the path optimized to manipulate the CO-molecule along the Cu-axis.

In conjunction with the path planning algorithms, the trained reinforcement learning (RL) model is employed to construct a unit cell of an artificial structure. **Figure 3** illustrates the outcomes of the structure construction process as the agent is fine-tuned during the deployment. Here, we seek to assemble a unit cell of an artificial graphene with lattice constant of distance 7*a* (where *a* ~ 2.5 Å is the lattice constant of the Cu(111) lattice). **Figure 3a** shows the evolution of the reward as a function of manipulation iterations and shows an increase in reward during the manipulation process, lasting until 120 iterations. **Figure 3b** presents the average episode reward and corresponding episode length as a function of the number of episodes. As the experimental iterations progress, we observe higher reward, and this is accompanied by shorter episode lengths.

In **Figure 3c**, the action parameters predicted by the model—namely, $V_S$, $I_t$, and $S$—are shown as sampled by the RL agent to guide the manipulation process. The predictions are skewed towards lower values of the $V_S$ and higher $I_t$, and lower $S$. We also see in **Figure S2** that the predictions are sampled as a distribution across the optimal values. **Figure 3d** displays the final constructed unit cell of an artificial graphene lattice, assembled through the autonomous manipulation procedure. The structural accuracy is governed by the reward tolerance parameter $\varepsilon$. In the structure shown, a value of $\varepsilon = 0.3$ nm is used, tolerating a placement error of up to one Cu lattice constant. Additionally, using the automated procedure, we formed a different structure of letters by arranging the CO molecules to form the "ORNL" structure shown in **Figure 3e**.

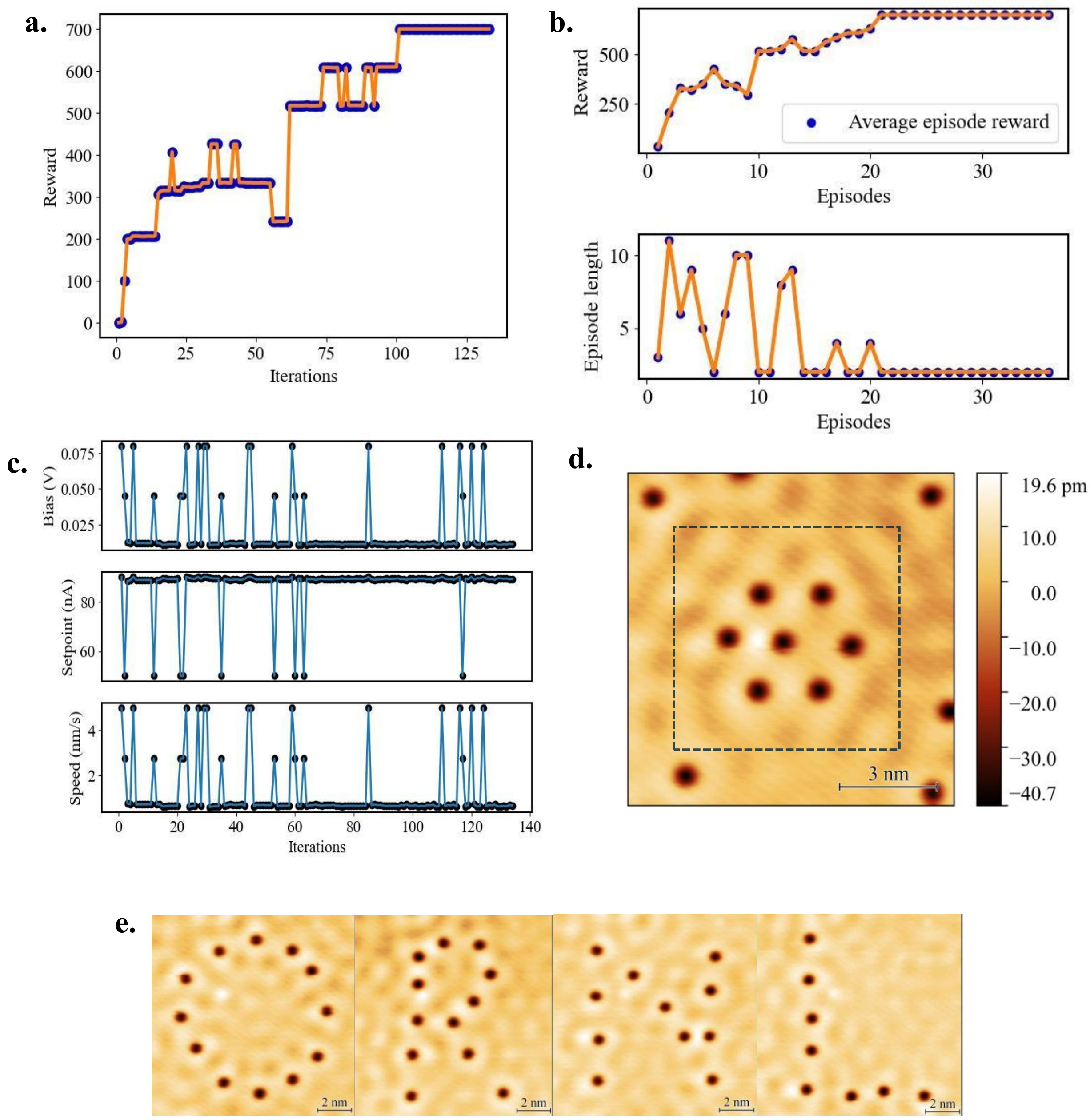

**Figure 3**: RL deployment and fine-tuning for structure creation. **(a)** shows the variation of reward as a function of manipulation iterations. **(b)** Relationship between the average episode reward and episode length with respect to the number of episodes. **(c)** Model-predicted action parameters of bias, setpoint and manipulation speed were sampled by the RL agent. **(d)** The final structure of a graphitic unit cell with a triangular lattice distance 7*a*, constructed using the autonomous manipulation procedure. **(e)** The "ORNL" structure was constructed using an RL-based manipulation procedure.

We conducted a comparative analysis to evaluate the performance of the RL agent in predicting effective action parameters against a baseline random sampling policy. **Figure S3** summarizes the results of this comparison, where the initial 100 manipulation iterations utilized randomly selected action parameters, followed by the deployment of the RL agent for the remaining sequence.

**Figure S3a** illustrates the reward trajectory as a function of manipulation iterations, with the blue-shaded region denoting the random sampling phase. **Figure S3b** presents the relationship between average episode reward and episode length over time, where the first 10 episodes correspond to random sampling, after which the RL agent is deployed. An eventual reduction of the episode length is observed after the random sampling regime, indicating efficient prediction using the RL agent. **Figure S3c** captures the progression of the action parameters—bias, setpoint, and manipulation speed—across iterations and shows the predictions shift from stochastic sampling to model-predicted optimal values.

Next, we implemented the construction of a larger graphitic artificial lattice, with the CO triangular lattice constant of distance 6*a*, consisting of 37 CO molecules, shown in **Figure 4**. The high-symmetry centers of the lattice form the targeted graphitic geometric pattern. It is to be noted that this experiment was performed with a new tip, and therefore, we retrained the model with an appended batch of training data consisting of manipulation procedures. This additional data (~ 2000 transitions) was integrated into the original dataset to ensure that the policy maintained generalization while simultaneously learning subtle characteristics of the new tip. The realized structure was constructed in multiple stages, snapshots of which are shown in **Figure 4a-d**. The structure shown in **Figure 4d** is the final constructed artificial lattice. We observed that at times, the tip state changes, and this was observed as a reduction of the manipulation efficiency across iterations. At times, we observed that the molecules repeatedly moved opposite to the manipulation direction, that is likely attributed to changes of tip apex. In such a condition, we had the human operator intervene, condition the tip, and reinitiated the automated manipulation procedure. Additionally, molecule positioning over short distances ($< 0.5$ nm) is dependent on the hyperparameters of the path-planning algorithms (details of hyperparameters in methods section). These parameters were optimized in the event of reduced manipulation efficiency. The reward profile across the entire manipulation process is shown in **Figure 4e**. The vertical blue lines indicate the iterations resumed after tip conditioning. The parameters that were predicted are shown in **Figure S4**. We particularly observe that the predicted speed values are higher than those obtained in the previous experiment, which is presumably attributable to the characteristics of the new tip.

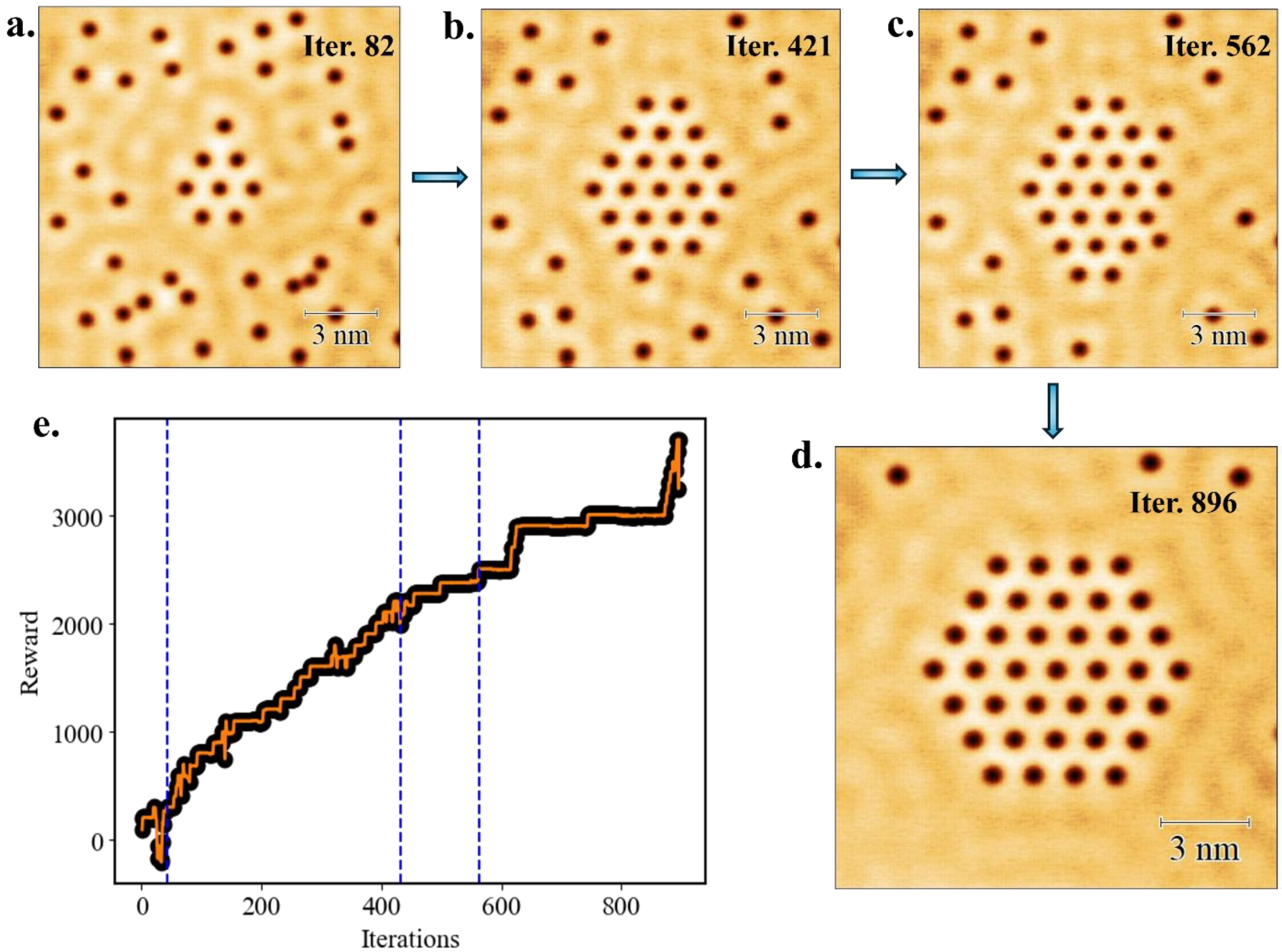

**Figure 4**: Construction of the artificial graphene lattice using the RL agent. Panels **(a)** – **(d)** depict the different stages of the assembled structure. **(e)** The reward value is shown across the iterations. The blue vertical line depicts the iteration at which the tip was conditioned, requiring operator intervention.

The constructed lattice was characterized using scanning tunneling spectroscopy (STS), and the observations are described in **Figure 5**. It is to be noted that the electronically active regions of the structure correspond to surface areas uncovered by CO adsorption, where nearly free electrons from the Cu(111) surface state reside. These "artificial atom" positions are indicated by green scatter points in the inset of **Figure 5a**. Each of these regions behaves as an artificial atom, collectively forming a honeycomb lattice. **Figure 5a** shows the normalized differential tunneling conductance (*dI*/*dV*) spectra acquired at the high symmetry points at the center of the triangular lattice (spectral acquisition sites are shown in the inset of **Figure 5a**). To isolate the spectroscopic signatures associated with the graphitic lattice, *dI/dV* spectra collected within the artificial lattice were normalized with respect to the spectrum acquired on the pristine Cu(111) surface (shown in **Figure S5**). The *dI/dV* spectrum shown in **Figure 5a** indicates the conductance minimum at ~ 80 mV, attributed to the Dirac point, along with the shoulder-like features corresponding to *M*-point in reciprocal space, marked as $E_M$. These features are consistent with the trends observed in previous reports.[3] **Figure 5b** which displays line-*dI/dV*-map, demonstrates the consistency of the Dirac point across the width of the artificial lattice. To unveil the spatial distribution of these features we collected *dI/dV* maps in constant height mode at the $V_S$ corresponding to the Dirac point (80 mV) and at $E_M$ (250 mV), shown in **Figures 5c** and **5d,** respectively. The *dI/dV* map acquired at $V_S$ = 80 mV (**Figure 5c)** reveals the reduced intensity over artificial graphene lattice, while the *dI/dV* map taken at $V_S$ = 250 mV (**Figure 5d**) highlights the enhanced intensity, in agreement with the spectral profile seen in **Figure 5a**

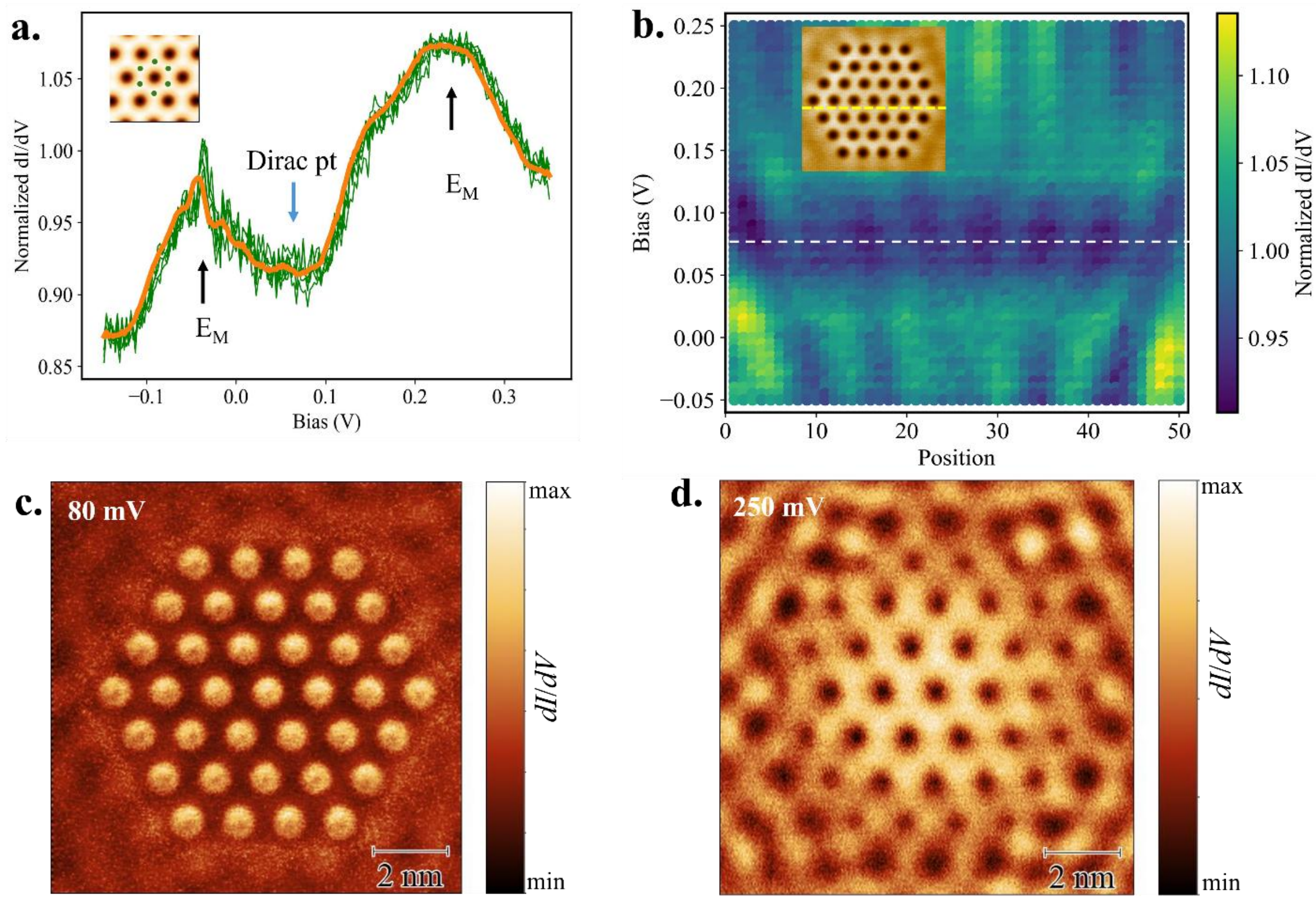

**Figure 5**: Electronic structure of the artificial lattice. **(a)** The normalized *dI/dV* spectra acquired in vicinity of the center of the artificial lattice. Each spectrum is normalized with the *dI/dV* spectrum of the pristine Cu (111) surface. The *dI/dV* acquisition sites which correspond to the "artificial atom" positions, are shown in the inset. **(b)** Normalized *dI/dV* spectra acquired along the line traversing artificial graphene shown in the inset. **(c)** constant height *dI*/*dV* maps taken at *Vs* of 80 mV, corresponding to the energy position of Dirac point. **(d)** *dI/dV* map collected at *Vs* 250 mV, corresponding to the energy $E_M$.

## Discussion

This work demonstrates the use of a deep reinforcement learning agent for atomic manipulation to create artificial lattices. While full autonomy still requires a complete robust tip conditioning procedure, this work demonstrates semi-automated approach reliant on human-guided inputs and operations. Moreover, we find that path planning benefits substantially; to avoid unnecessary interactions with pre-existing adatoms/molecules on the surface and the knowledge of the crystallographic axes is utilized to minimize energy of the manipulation path in this case. One of the bottlenecks in extending manipulation procedures for larger structures lies in scalability. In our experiments, each manipulation iteration required approximately 1-1.5 minutes. The 37-molecule system required approximately 900 iterations (as shown in **Figure 4e**) resulting in a total construction time of 25 - 30 hours (this includes tip-conditioning procedures). Although the process is largely autonomous and does not require the presence of the human operator, extended durations inevitably introduce tip instabilities and state changes. To address these constraints, future work will focus on strategies to reduce time-per-iteration using sparse-scanning protocols and live-tracking algorithms based on z-signal trajectories.

Moreover, automated tip conditioning is expected to be an integral part of the manipulation capability. In practice, tip conditioning is specific to the experimental modality – be it imaging, spectroscopy or atomic manipulation. Based on previous reports,[24, 31] the tip conditioning routines can be extended to frameworks that incorporate manipulation efficiency-metrics into the reward function. This, in conjunction with tip-shaping routines, can be incorporated within the experimental loop. Therefore, further efforts should seek to combine the elements of tip-conditioning methods, state-of-the-art path planning methods combined with property optimization routines.[38] Alternately human expertise can be incorporated to better train agents, either in the form of specific structured priors, a low number of practical demonstrations, or both. Most importantly, in this study, we have demonstrated autonomous manipulation of CO to generate artificial graphene lattice with characteristic Dirac points within the electronic configuration.

However, of more interest is the specific inverse problem of designing the artificial lattice structure given the electronic state of interest. A variety of ML-based methods, combined with theoretical calculations, are the obvious next step in combining these RL agents towards realizing the inverse design of artificial lattices for designer quantum states. Genetic algorithms or generative ML models[39, 40] trained on large-scale *ab initio* calculations can be used to predict novel artificial nanostructures with exotic topological properties that are hard to achieve in natural materials. Edges or interfaces between topological 2D structures are expected to host topologically protected edge-states due to the bulk-boundary correspondence principle. Specifically, breaking time-reversal symmetry using magnetic adatoms instead of CO molecules and using superconducting substrates such as Nb or $Nb_2Se_3$ instead of Cu(111) can help us to design atomically precise topological qubits,[41] which can now be placed at will to form interesting topological assemblies. Forming other types of lattices such as Kagome lattice that can give rise to flat-bands, geometric frustration as well as Dirac bands, while at the same time using the STM tip to control filling of these states can open a robust approach to interrogate a host of emergent physics that are difficult to probe in natural materials due to defects and hidden interfaces. As such, we expect that our approach to automate such artificial 2D geometric structure will accelerate realization and new breakthroughs in understanding such emergent quantum states. Furthermore, using inverse learning[42] approaches, one can also directly extract interactions of a Hamiltonian from spectra, allowing a unique testbed to benchmark many-body theoretical methods, which has never been possible before.

**Conclusion**

In this work, we present a methodology for automated molecular manipulation using a scanning tunneling microscope (STM), along with its application to the construction of artificial structures. The automated workflow combines object detection to localize individual molecules and reinforcement learning to predict optimal manipulation parameters. In our implementation, the model is first trained offline and subsequently fine-tuned online during deployment. Additionally, path-planning algorithms are incorporated to enhance manipulation efficiency, particularly in scenarios where molecules seem immobilized. Using this approach, we constructed an artificial graphene lattice and performed its characterization.

**Code Availability**:

The code used for the RL and the automated experiments are available at: github.com/gnganesh99/RL_STM_d3rlpy.

**Acknowledgement**

The experimental research was supported by the U.S. Department of Energy, Office of Basic Energy Sciences, Scientific User Facilities Division as part of the QIS Infrastructure Project (FWP ERKCZ62), "Precision Atomic Assembly for Quantum Information Science" and performed at the Center for Nanophase Materials Sciences (CNMS). The algorithmic development was supported by the Center for Nanophase Materials Sciences (CNMS), which is a US Department of Energy, Office of Science User Facility at Oak Ridge National Laboratory.

**Conflict of Interest**

The authors declare no conflict of interest.

**Author Contribution**

†GN and MT contributed equally to this work. GN prepared the code for offline training and for the automated experimental workflow based on inputs from RV. MT prepared the experimental setup. GN and MT performed the manipulation experiments and compiled the data. WY performed the spectroscopic studies and assisted in designing the object detection workflow. APB, APL, and RV conceived the idea, supervised the implementation of the project, and assisted with data

interpretation. PG contributed to the conception of the original idea. All authors contributed to the preparation of the manuscript.

**Supporting Information**

# Automated Construction of Artificial Lattice Structures with Designer Electronic States

Ganesh Narasimha[1], Mykola Telychko[1], Wooin Yang[1], Arthur P. Baddorf[1], P. Ganesh[1], An-Ping Li[1], Rama Vasudevan[1]*

[1]*Center for Nanophase Material Sciences (CNMS), Oak Ridge National Laboratory (ORNL), Oak Ridge, Tennessee, USA – 37831*

* *vasudevanrk@ornl.gov*

## Table of Contents

**Loss function – RL offline training**

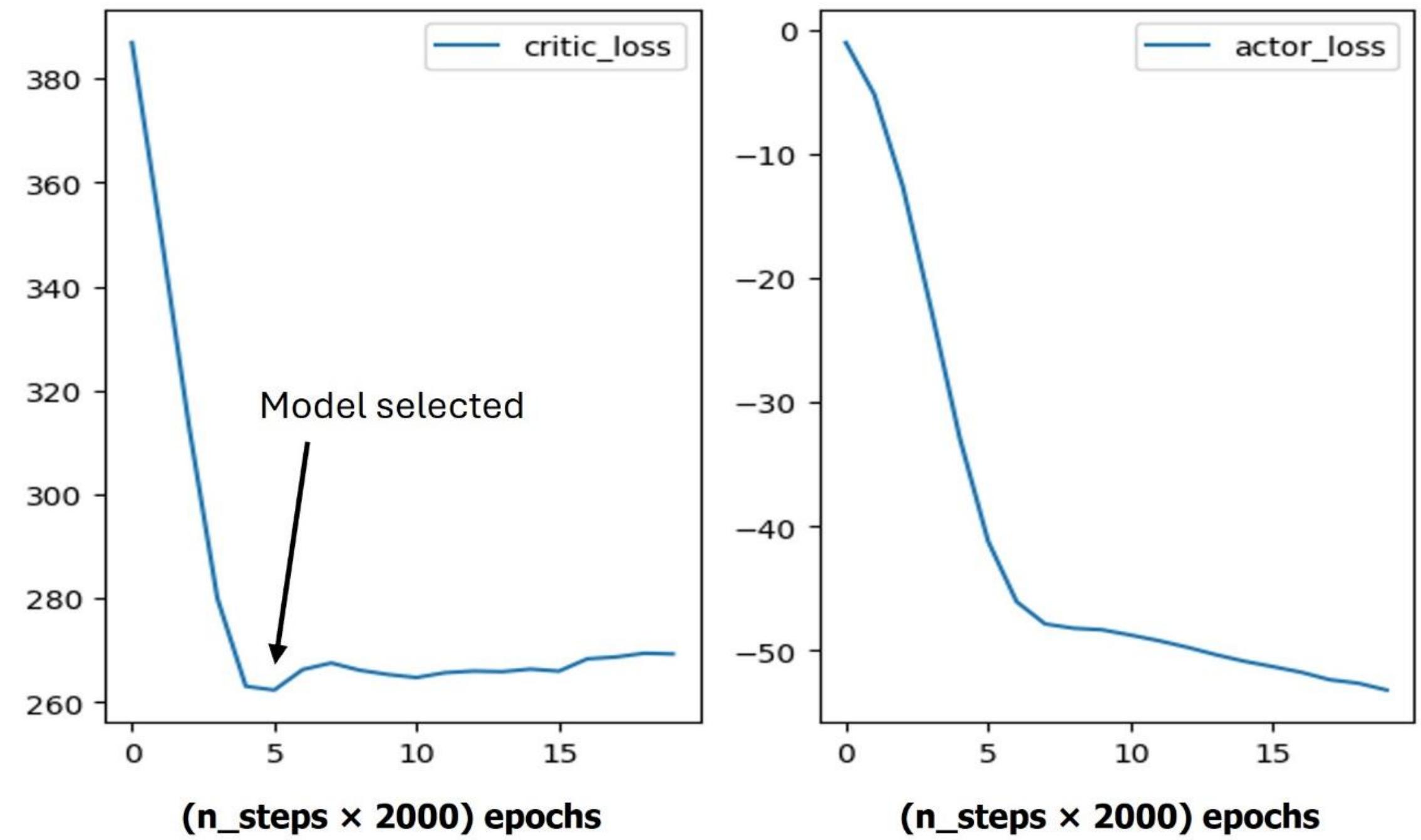

Figure S1: Decay of the loss value of the actor and critic network during offline training.

**Distribution of predicted parameters during online deployment**

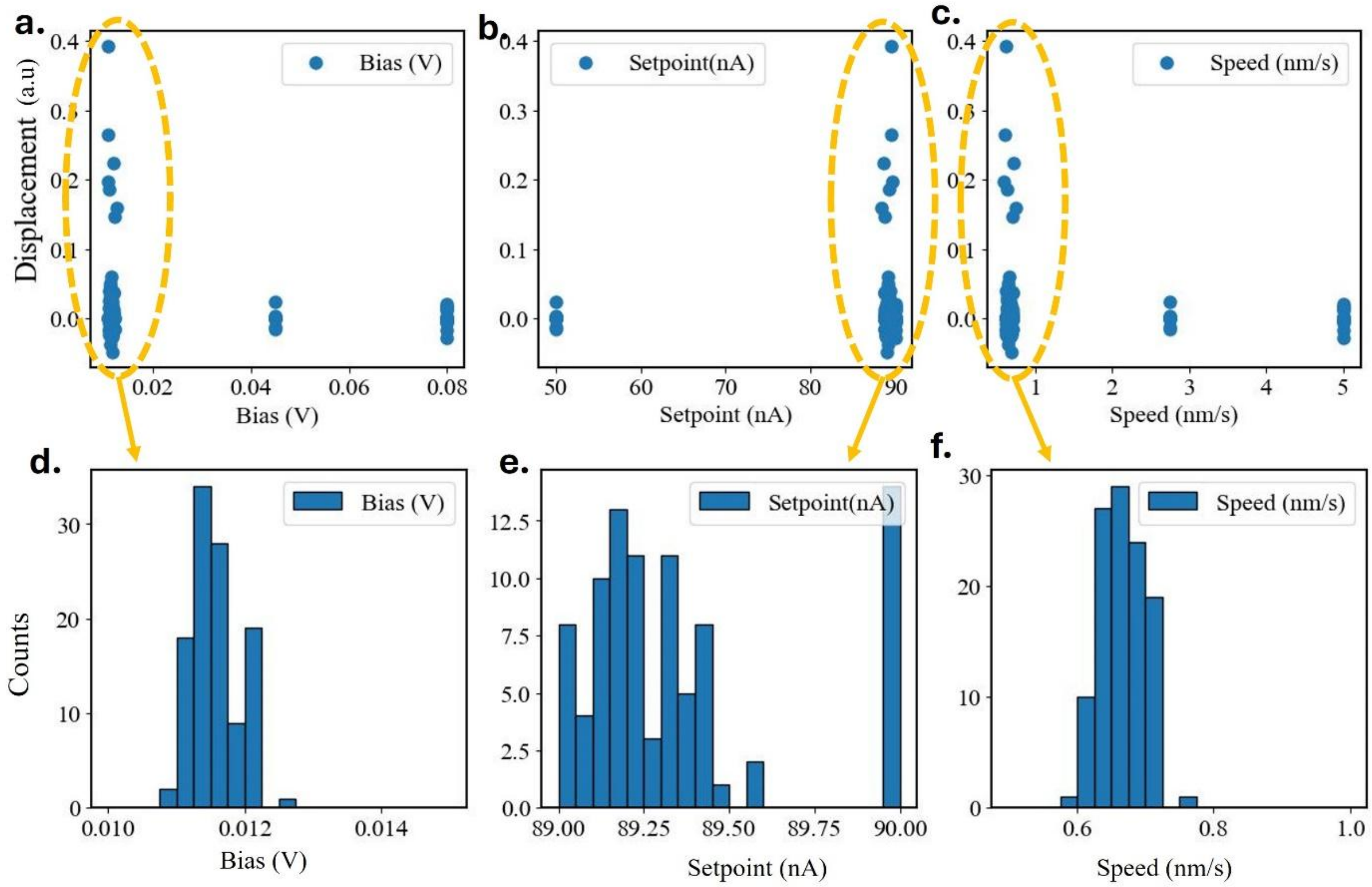

**Figure S2**: Parameters prediction during the construction of the artificial structure shown in Figure 3d.

**Comparison of RL predictions with random policy**

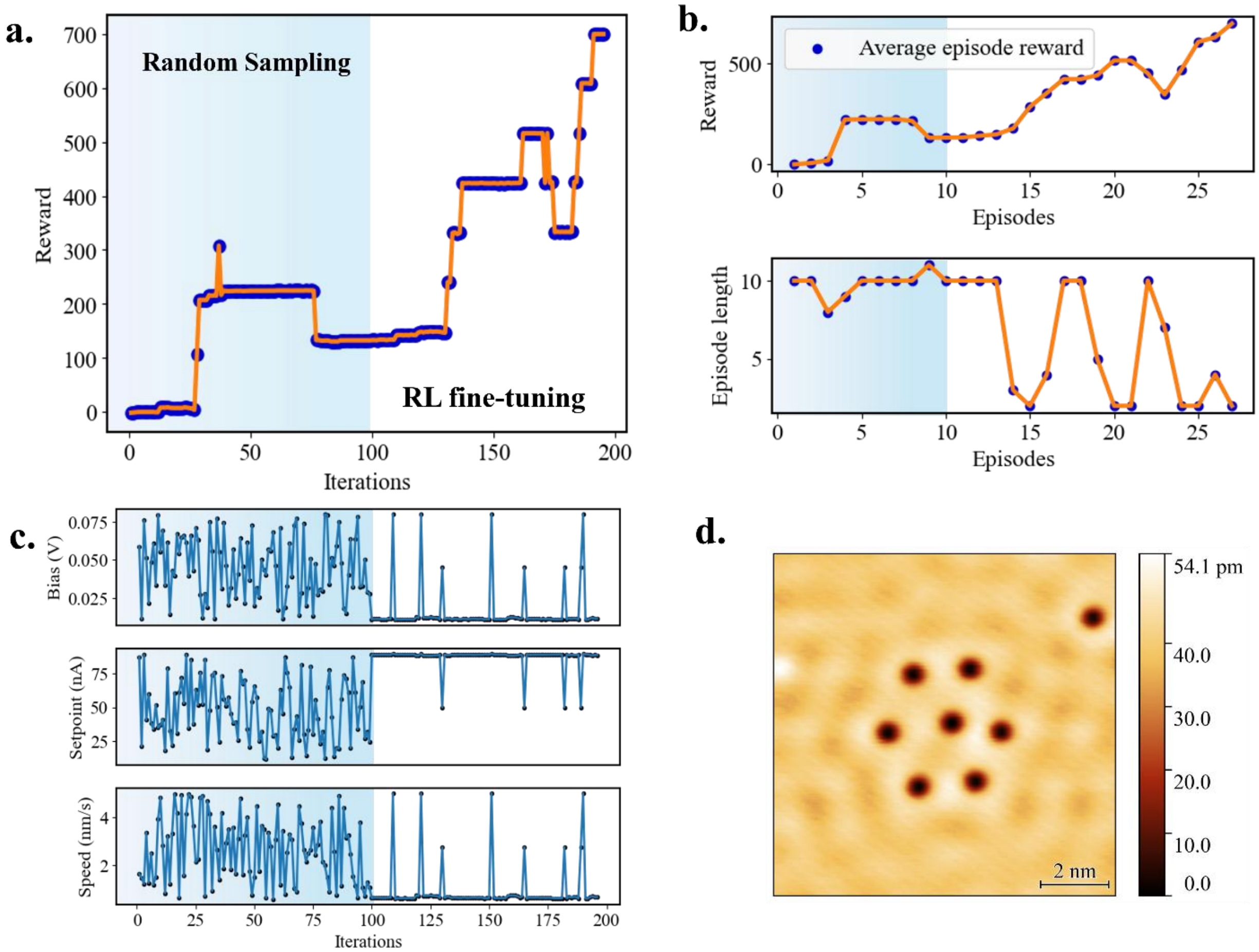

**Figure S3**: Comparison to random sampling **(a)** shows the variation of reward as a function of manipulation iterations. The blue-shaded region corresponds to the initial 100 iterations where the action parameters are randomly sampled. **(b)** Shows the comparative relationship between the average episode reward and episode length with respect to the number of episodes. The initial 10 episodes correspond to episodes where the action parameters are randomly sampled. **(c)**Variation of the sampled action parameters across the iterations. The initial 100 action parameters (in the blue shaded region) are randomly sampled followed by the model-predicted action parameters. **(d)** Final structure constructed using the automated manipulation procedure.

**Parameter predictions - construction of the graphitic lattice.**

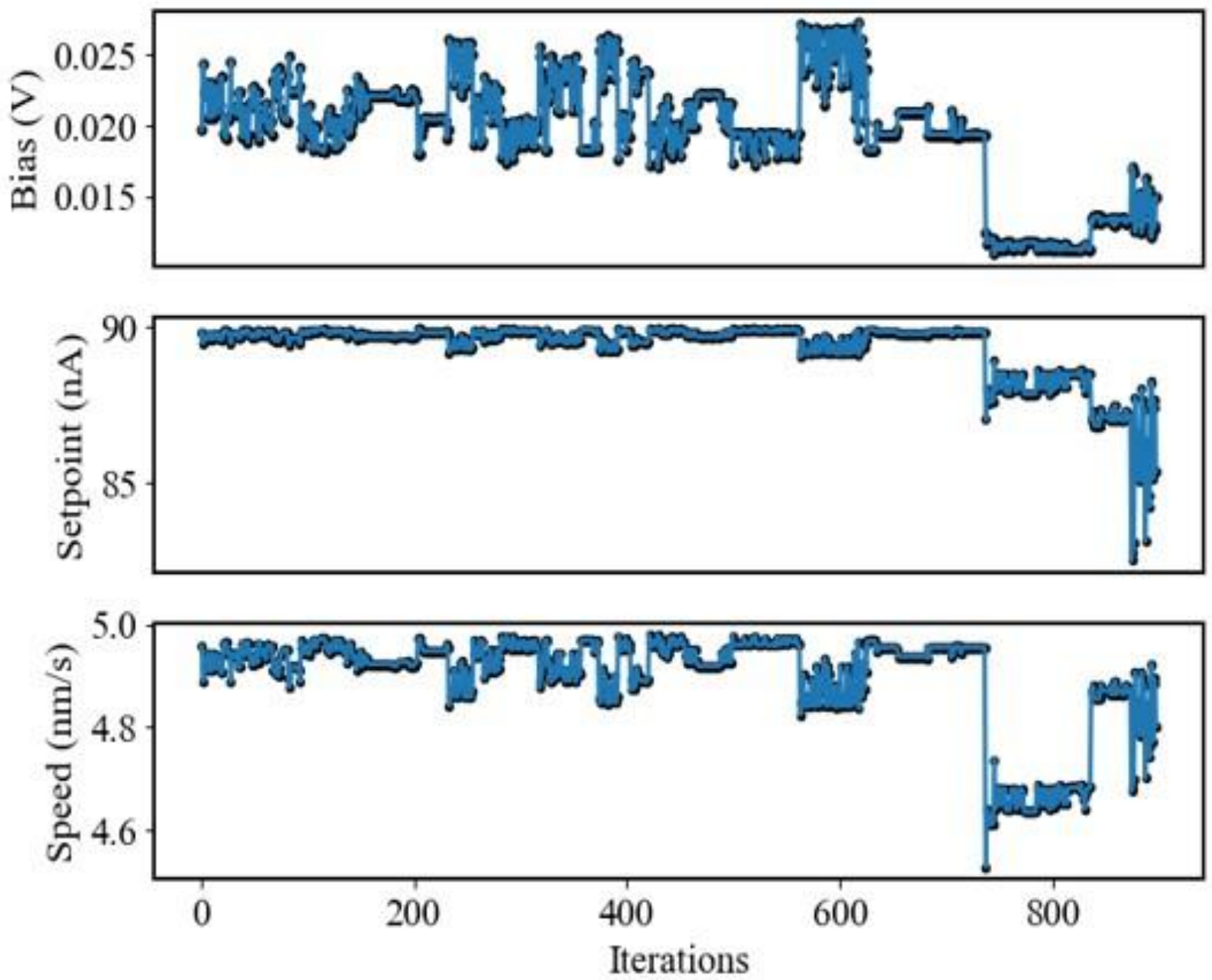

**Figure S4**: Parameter predictions across the iterations during the construction of the graphitic lattice

**Conductance spectrum on pristine Cu surface**

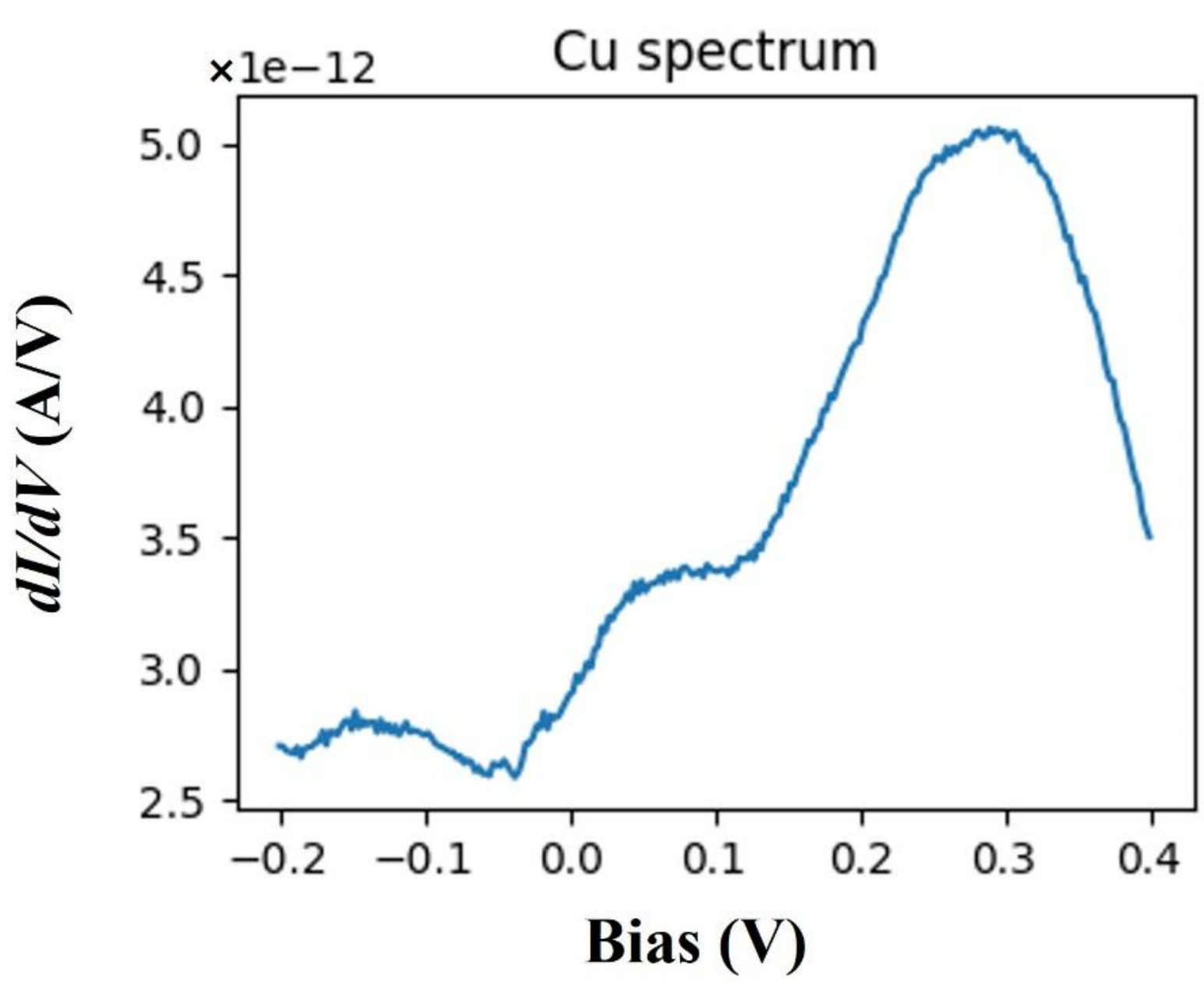

**Figure S5**: Differential conductance spectrum of the pristine Cu surface.

## Methods

I. STM experimentation

STM experiments were carried out under UHV conditions at temperature 9.7K using a commercial Omicron Infinity SPM instrument. The Cu(111) single crystal (MaTeck *GmbH*) was routinely cleaned by multiple cycles of $Ne^+$ sputtering and annealing. The CO gas was dosed using a leak-valve (pressure $1x10^{-8}$ mbar) for 10 seconds, while Cu(111) was held at ~12 K. The *dI/dV* data were acquired using lock-in amplifier using modulation frequency of 743 Hz and modulation voltage of 10 mV.

While we assume the lattice constant of the Cu lattice to be ~ 2.54 Å, minor variations due to the calibration of the piezo scanner can introduce errors into the final positioning, especially as we scale the structure. To account for this, we first manually arranged one unit cell of the graphitic lattice and recalibrated the piezo-imaged lattice constant as 2.26 Å. We then used the corrected value to re-design our graphitic lattice before commencing the automated manipulation process all over again.

II. Reinforcement Learning

The reinforcement learning (RL) methods were incorporated using the d3rlpy package (https://github.com/takuseno/d3rlpy) that contains the code base for using the RL models.[35] Here we used the DDPG (deep deterministic policy gradient) model for training the data.[36, 43] DDPG uses an actor-critic network. These networks consist of MLPs that have two hidden layers with 256 neurons each, with *tanh* activation.

The training data was augmented by introducing transformations to the original buffer. The transformations included adding Gaussian noise to the state variables, with a noise factor in the range [0.01, 0.05]. We also added a small noise (noise factor = 0.001) to the action space to increase variability. This was also accompanied by shuffling the episodes.

The hyperparameters used for training and during the fine-tuning are as follows:

| **Hyperparameter** | **Value** |
|---|---|
| **actor learning rate** | 1e-6 |
| **critic learning rate** | 5e-6 |
| **optimizer** | Adam |
| **gamma** | 0.8 |
| **tau** | 0.005 |
| **epsilon** | varied in range [0, 0.1] |

**Table S1**: Hyperparameter values used for the RL model training and fine-tuning

From the images, we obtained the following observation variables: current position of the molecule ($X_{current}$), target position ($X_{target}$), and the manipulation angle($\theta$). These were input as state variables normalized in the range [0, 1]. The action variables of bias, setpoint, and tip speed were normalized in the range [-1, 1]. During the prediction, we observed exploratory predictions beyond the indicated range. However, these predictions were clipped to the indicated range.

## III. Alignment of Cu-lattice orientation.

We had the Copper with Cu (111) surface termination, onto which we carried out the manipulation experiments. The determination of the Cu (111) axis is crucial for the precise positioning of the molecules since the CO molecule positioning is energetically favorable to be located on the Cu-site.

We begin by conjoining three CO molecules, such that the distance between two consecutive molecules is spaced at the distance of 2a (where a=2.54 Å is the lattice constant of the Cu (111) surface). This condition is satisfied when the molecules are axially attached along the Cu-axis. Following this, the center molecule is displaced, and a similar approach is implemented to find the other Cu-axis. Once we find the two primary axes, the scan frame is rotated to horizontally align with the axis of Cu.

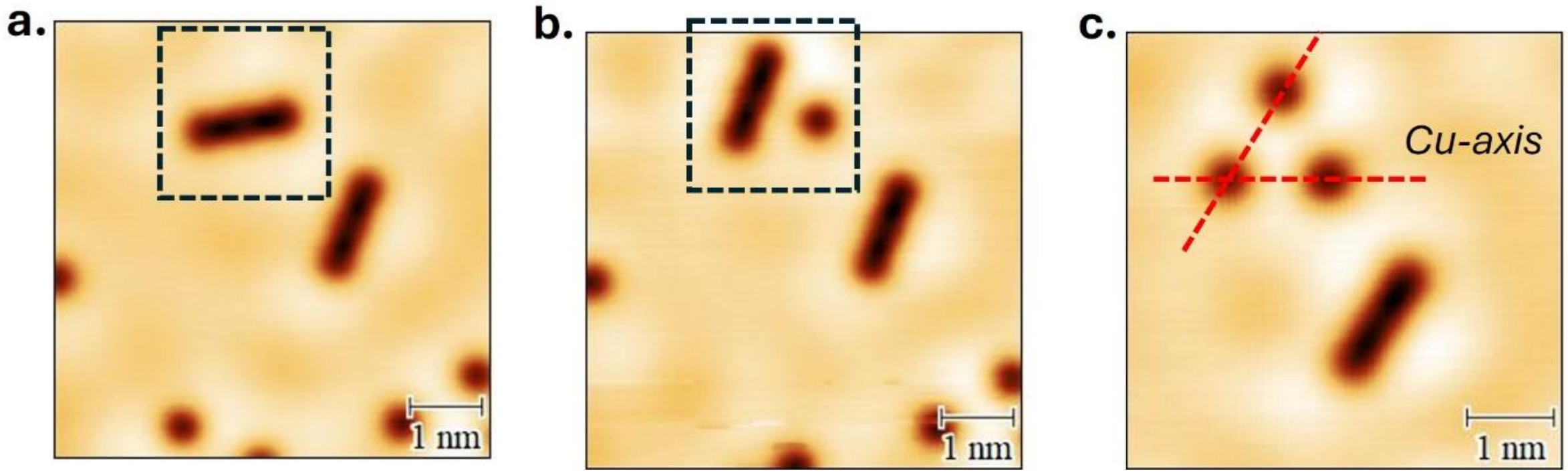

**Figure S6**: Procedure for determining and aligning the Cu (111) axis. **(a)** Conjoining CO molecules in one axis that corresponds to the Cu lattice. **(b)** A similar attachment of the molecules in the other Cu-axis. **(c)** Center atoms were displaced along the two axes. The scan frame is rotated to align horizontally with the Cu-lattice.

IV. Molecule Detection using YOLO

For the object detection of CO molecules, we used the YOLO method. We experimented with two kinds of model – yolov8s and yolov10s, from ultralytics.[34, 44] While the yolov8s is reliable for single molecule detection, we found the detections to be more efficient in yolov10s to resolve joint CO molecules. We deployed the trained yolov10s model for the construction of artificial lattice. For training, we used ~ 100 annotated images of the CO molecules and augmented the training dataset with image transformations.

In the experiment, the confidence threshold was set to a lower value of 0.2 to identify CO molecules under instances of poor image resolution. The intersection over union (iou)-threshold was set to an intermediate value of 0.5 to identify joint molecules that appear as overlapping objects. Some examples of the CO molecule detections are shown below.

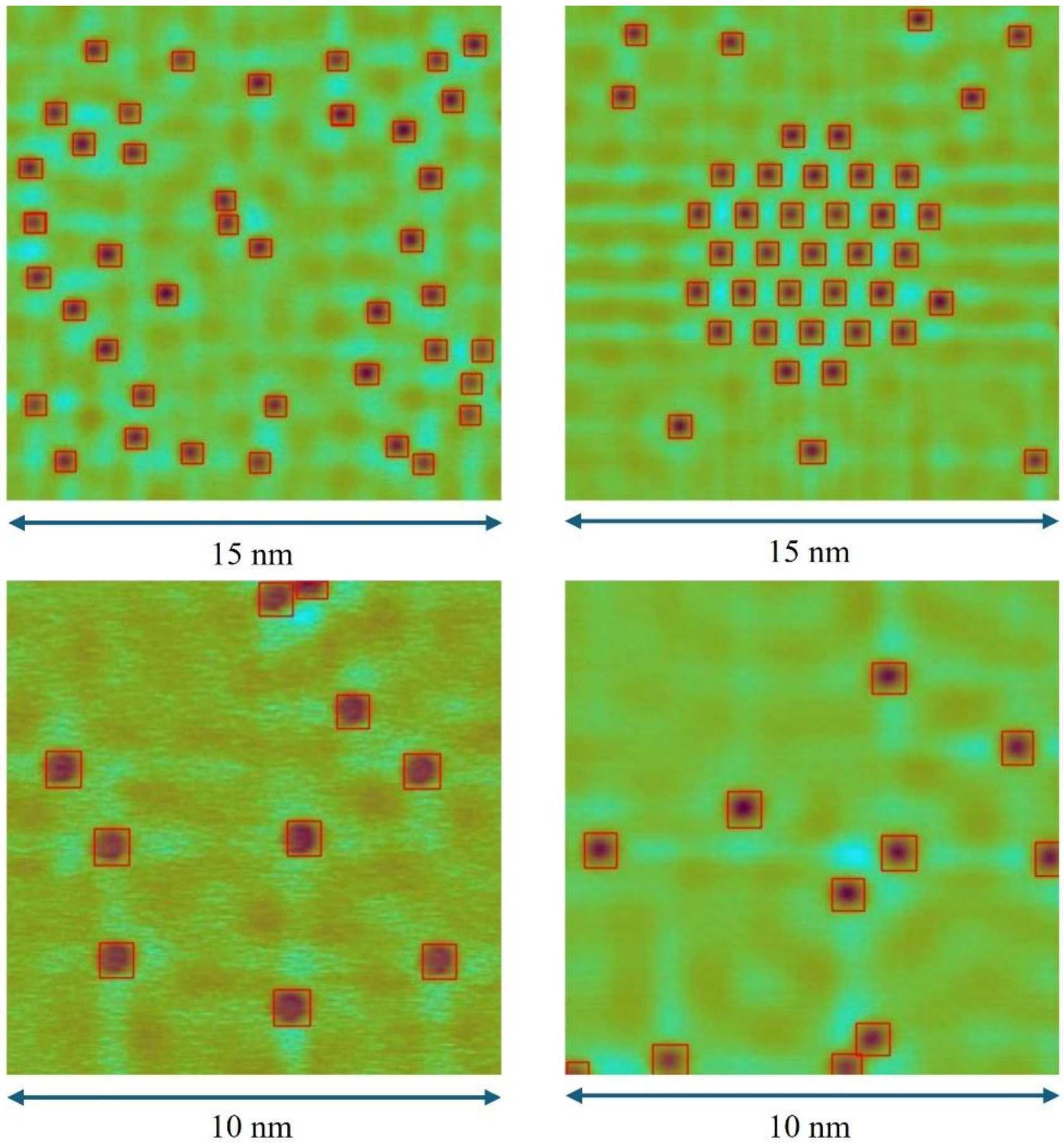

**Figure S7**: Examples of the YOLO based method for the detection of CO molecules. The red bounding boxes indicate the CO detections.

## V. Linear Assignment

We used a linear assignment algorithm available on the *scipy* library for to assign molecule position to the target, based on the distance cost matrix. As per *scipy* documentation, this function uses the modified Jonker-Volgenant algorithm.[45] Our code also describes an additional distance-collision cost function that introduces additional cost if the molecule path encounters a possible collision. However, we found that the distance-cost matrix works for simple structures.

We encountered another bottleneck, especially while dealing with numerous molecular configurations (> 10) – the molecule-target assignment was not consistent throughout the whole manipulation procedure. To solve this problem, once the initial assignment is completed, in every iteration, we freeze the assignment of molecules currently not considered for manipulation – thereby preserving the assignment of the moving molecule to the designated target. Examples of multi-molecular assignment are shown below.

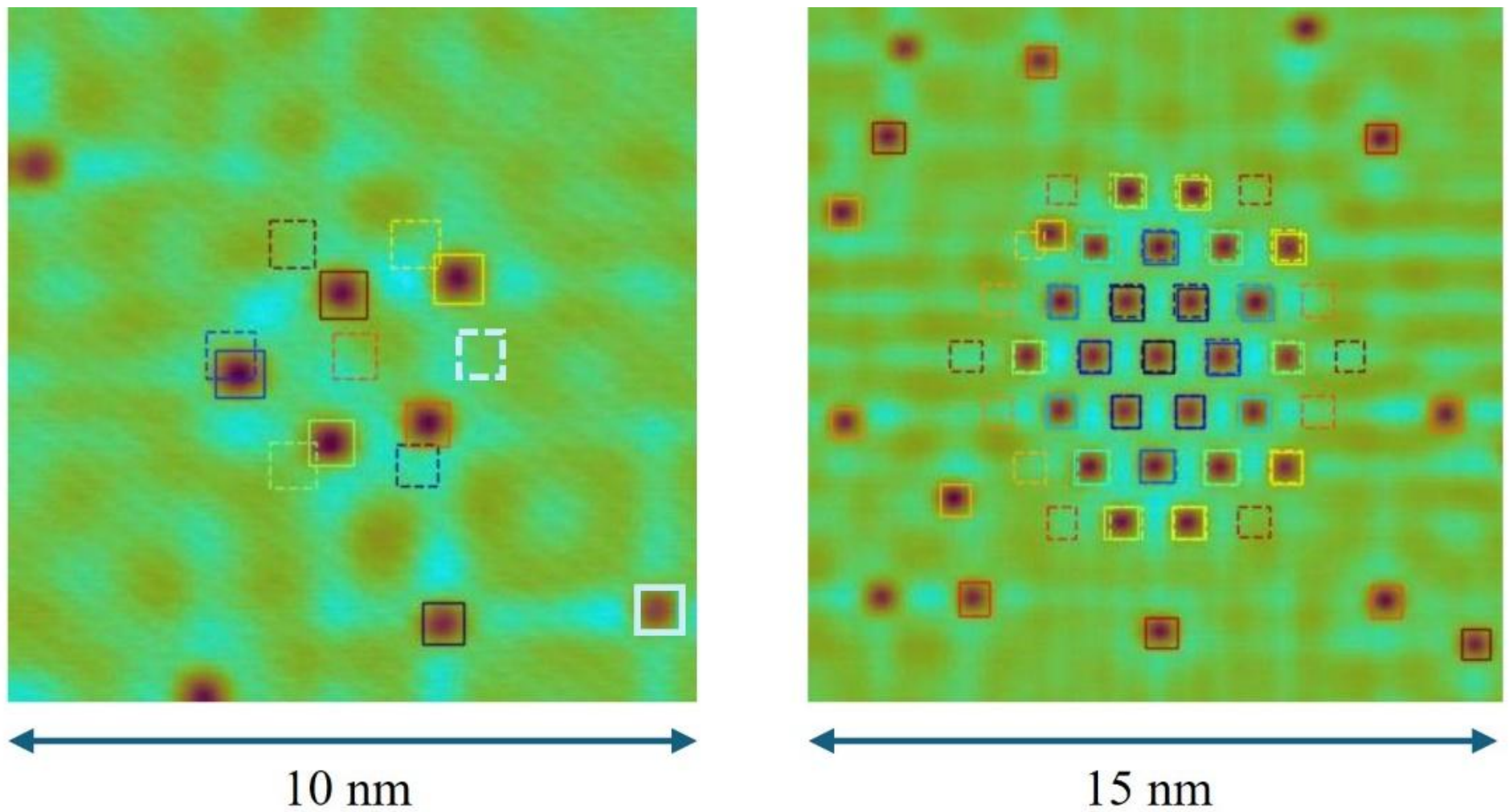

**Figure S8:** Linear assignment of the molecules to the target positions. The solid-lined boxes indicate the molecule detections while the similarly colored dash-lined boxes indicate the assigned targets.

## VI. Path Planning

As introduced previously in the manuscript, these path-planning methods are used to facilitate the movement of stuck molecules that seem immobile.

- Target Offset

Target offset is introduced to overmanipulate the molecule beyond the target as shown in Figure S10. A molecule is denoted as “stuck” when the position of the molecule (that is considered for manipulation) doesn’t change across the iterations. The offset positioning increases by a factor of 0.1×molecule_width (molecule_width is derived from the size of the detected bounding box). This target offset increases up to a certain threshold, after which we modify the manipulation path to align with the Cu-axis.

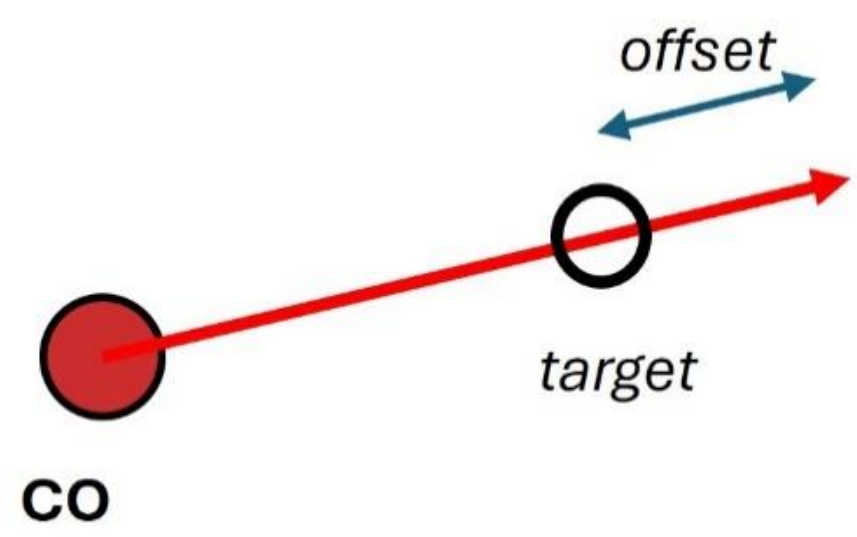

**Figure S9**: Illustration of the target offset method for molecule positioning

- Manipulating along Cu-path

This procedure attempts to optimize the manipulation path by performing incremental manipulations along the Cu-axis.  first determine the manipulation angle from the molecule to the target position. Then, we choose the manipulation angle closest to the Cu axis orientation (0°, 60°, 120°, 180°, 240°, 300°, 360°) and perform incremental manipulations of 0.1 × manipulation distance.

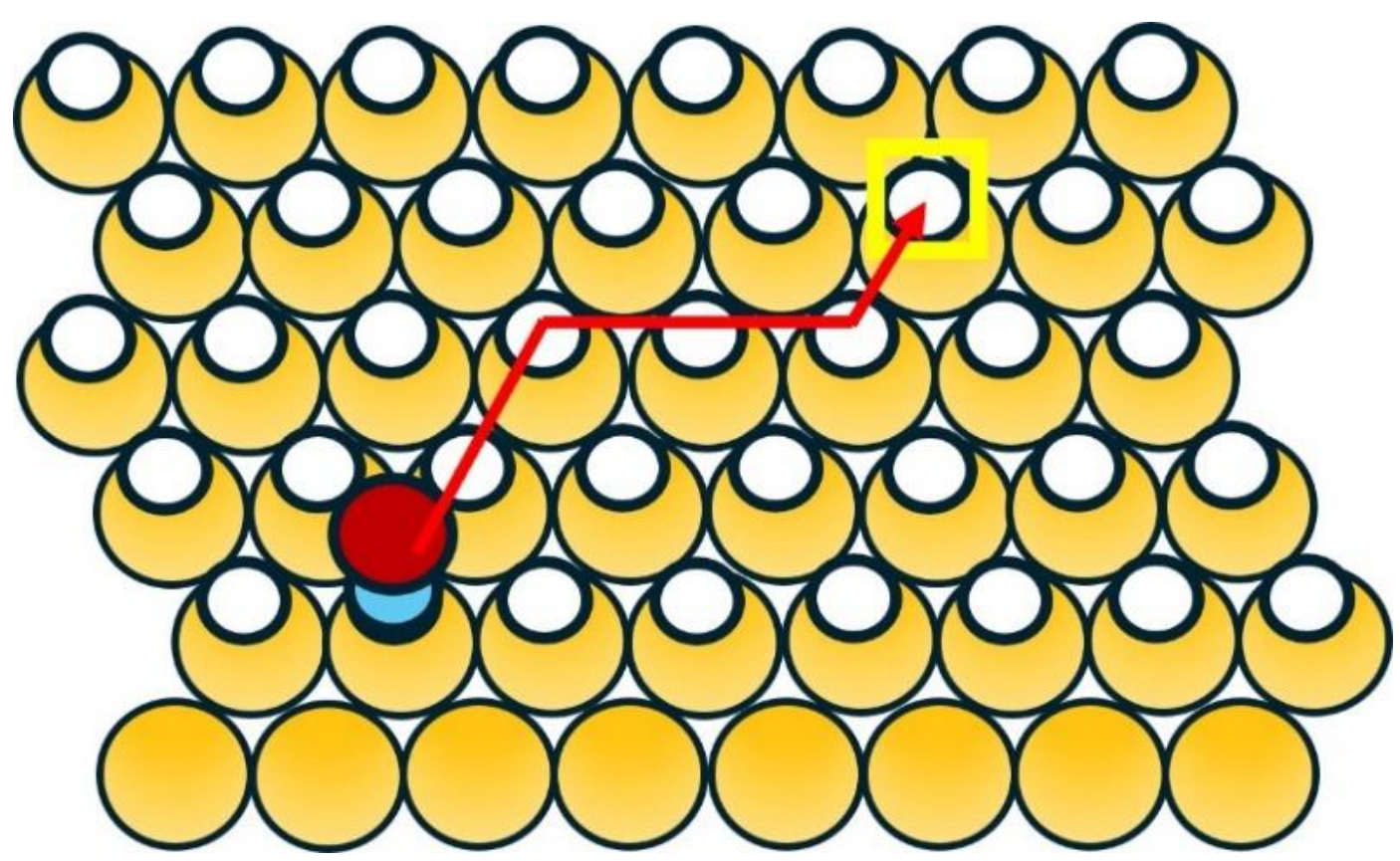

**Figure S10**: Depiction of the path optimized to manipulate the CO-molecule along the Cu-axis.

This procedure is repeated a few times until a set limit is reached, following which the workflow performs random incremental manipulations along the Cu-axis.

## VII. Drift correction

We implemented drift correction to (i) accurately determine manipulation distance by correcting the drift-induced component and (ii) compensate the drift by repositioning the scan frame.

To determine the drift, we define an anchor molecule whose position is considered fixed and acts as a fiducial marker within the scan frame. This is either an extra molecule that is not part of the manipulation set or is the central molecule of the target configuration. At every iteration, the program dynamically determines the deviation of the anchor molecule from the designated target and repositions the scan frame to compensate for the drift.

Although additional mechanisms such as the averaged drift of static molecules and phase-cross correlation functions are presented within the code, we find the above-described method is reliable especially when we have a sparse distribution of the molecules.

During long-term manipulations, we encounter events where tip change results in a drastic variation of the images, sometimes resulting in noisy features. In such a scenario, erroneous detections of molecule positions can result in overestimation of the drift value, which can result in an uncontrolled movement of the tip across the surface. To avoid these accidents, we have set a safety upper threshold (1.5×average_drift) to the determined drift value, beyond which, the tip and the scan frame position remain unaltered.

## VIII. Automated STM controls

To enable external control of the STM, the Nanonis software offers a TCP-based interface that allows third-party applications to send and receive commands via a dedicated service port. This interface supports the transmission of low-level byte streams, which the software interprets as operational instructions for STM manipulation. The communication protocol and API required for this interaction are provided by Nanonis. Our experimental framework incorporates an STM communication module, developed with reference to an open-source library: https://pypi.org/project/nanonisTCP/[33]